\relax
\documentclass[letterpaper]{article} 
\usepackage{aaai21}  
\usepackage{times}  
\usepackage{helvet} 
\usepackage{courier}  
\usepackage[hyphens]{url}  
\usepackage{graphicx} 
\usepackage{natbib}  
\usepackage[labelfont=bf]{caption} 
\usepackage{booktabs}
\newcommand{\xhdr}[1]{\vspace{1.7mm}\noindent{{\bf #1.}}}

\usepackage{multirow}

\usepackage{amsmath}
\usepackage{amssymb}
\usepackage{makecell}
\usepackage{float}
\usepackage{subcaption}
\usepackage[hidelinks]{hyperref}
\usepackage{placeins}

\usepackage[english]{babel}
\usepackage{csquotes}
\usepackage{xcolor}
\usepackage{verbatim}
\usepackage{booktabs}
\usepackage{breqn}
\usepackage[toc,page]{appendix}

\hypersetup{pdfinfo={
Title={Tube2Vec: Social and Semantic Embeddings of YouTube Channels},
Author={Léopaul Boesinger, Manoel Horta Ribeiro, Veniamin Veselovsky, Robert West}
}}

 \usepackage{booktabs}
\usepackage{caption}
\usepackage{subcaption}
\title{Tube2Vec: Social and Semantic Embeddings of YouTube Channels}
\author{
Léopaul Boesinger, Manoel Horta Ribeiro, Veniamin Veselovsky, Robert West\\
}
\affiliations{
EPFL\\
leopaul.boesinger@epfl.ch, manoel.hortaribeiro@epfl.ch, veniamin.veselovsky@epfl.ch, robert.west@epfl.ch
}

\begin{document}

\maketitle

\begin{abstract}
Research using YouTube data often explores social and semantic dimensions of channels and videos.
Typically, analyses rely on laborious manual annotation of content and content creators, often found by low-recall methods such as keyword search.
Here, we explore an alternative approach, using latent representations (embeddings) obtained via machine learning.
Using a large dataset of YouTube links shared on Reddit; we create embeddings that capture social sharing behavior, video metadata (title, description, etc.), and YouTube's video recommendations. 
We evaluate these embeddings using crowdsourcing and existing datasets, finding that recommendation embeddings excel at capturing both social and semantic dimensions, although social-sharing embeddings better correlate with existing partisan scores.
We share embeddings capturing the social and semantic dimensions of 44,000 YouTube channels for the benefit of future research on YouTube: \url{https://github.com/epfl-dlab/youtube-embeddings}.
\end{abstract}

\section{Introduction}\label{sec:intro}

Consider the following three YouTube channels:
\begin{itemize}
    \item[\textbf{A}] \textbf{Guns \& Gadgets}: (...) keeping YOU up to date on the constant attempts to infringe on the 2nd Amendment (...)
    \item[\textbf{B}] \textbf{Coalition to Stop Gun Violence}:  (...) CSGV seeks to secure freedom from gun violence through research (...)
    \item[\textbf{C}] \textbf{Mic the Vegan} --- Mic the Vegan is a vegan science writer that covers a variety of topics (...)
\end{itemize}
Semantically, \textbf{A} is more similar to \textbf{B} than to \textbf{C}, as both \textbf{A} and \textbf{B} talk about guns.
However, considering  the Left--Right spectrum (in the US context), \textbf{B} is more similar to \textbf{C} than to \textbf{A}, as both positions they support -- veganism and gun control -- are more prevalent within the political left, whereas unlimited restrictions towards gun ownership, supported by \textbf{A}, is a position associated with the political right.

Content's semantics and social dimensions (social constructs projected across a linear scale) greatly concern research using (or about) online platforms.
Researchers are sometimes interested in studying a specific topic and use heuristics to find semantically similar videos, tweets, or posts.  
For instance, on YouTube, a vast body of research has assessed the quality of medical information available on the platform~\cite{madathil_healthcare_2015}, where relevant videos and channels are usually found using the platform's own search engine.
Also frequently, researchers investigate the social dimensions of content present in online platforms, for instance, studying whether recommendations provided by YouTube are politically biased~\cite{hosseinmardi2021examining}, or age appropriate~\cite{papadamou2020disturbed}. 
Here, researchers rely on lists that manually place channels along the desired social dimension [e.g., the lists of left- vs. right-wing channels used by \citet{hosseinmardi2021examining}] or on classifiers trained on hand-labeled data [e.g., \cite{papadamou2020disturbed}'s deep learning method to detect videos that were not age appropriate for toddlers].
These approaches share two limitations.
First,  they are labor intensive, e.g., one would need to annotate a great many YouTube channels or videos manually.
Second, they may have a low recall, as it is hard to ensure that the heuristics used to find the data have captured a substantial portion of the relevant content in the platform. The latter shortcoming is particularly troublesome, as any heuristic may encode its own bias into the results threatening the validity of the downstream results.

Here, we explore an alternative approach to capturing YouTube channels' semantics and social dimensions. 
Instead of manually curating content or content creators, we let the data `speak for itself,' using latent representations derived from the data.
We use this approach to capture the social dimensions and the semantic similarity of 7.5M YouTube channels whose links were shared on Reddit between 2010 and 2022.
We create
1)~\textit{social sharing embeddings} by extrapolating previous work quantifying social dimensions on Reddit~\cite{waller_quantifying_2021};
2)~\textit{content embeddings} with the metadata associated with YouTube channels; and 
3)~\textit{recommendation embeddings} using the video suggestions given by YouTube. 
We validate these different embeddings on their ability to capture the social dimensions and semantics of YouTube channels using existing datasets and crowdsourced human ratings.
We find that recommendation embeddings excel at capturing both social dimensions and semantics, although social sharing embeddings better correlate with existing partisan scores.

\begin{figure}
    \centering
    \includegraphics{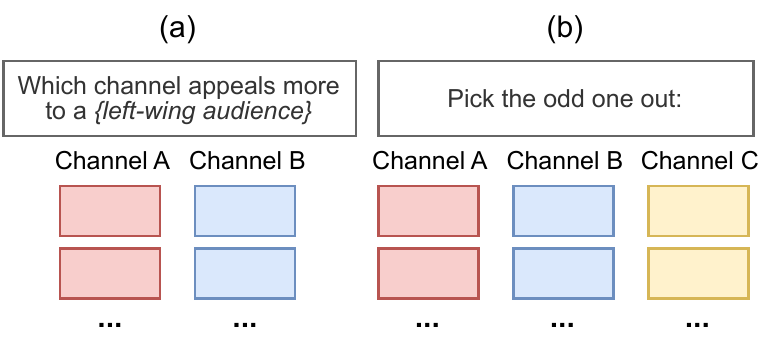}
    \caption{Depiction of the crowdsourcing tasks used to validate our embeddings. In (a), crowdworkers were asked to order channels in a specific social dimension; in (b), they were asked to select the dissimilar channel.}
    \label{fig:mturk}
\end{figure}

\section{Data}\label{sec:data}

\xhdr{Reddit data}
We used the Reddit Pushshift dataset~\cite{baumgartner_pushshift_2020},%
which contains nearly all posts and comments made on Reddit. 
Considering data between January 2010 and August 2022, we extracted tuples of $\langle$subreddit, video-id, author-id$\rangle$ for posts and comments with YouTube video URLs.
We then used YouTube's Data API to obtain the channel identifier for each video and relevant metadata for each channel (number of views, number of subscribers, creation date).
To link banned or removed YouTube videos to their respective channels, we used the datasets collected by \citet{ledwich_algorithmic_2019} and \citet{ribeiro_youniverse_2021} to re-hydrate the video identifiers of deleted channels.
Overall, we extracted 77.4M $\langle$subreddit, video-id, author-id$\rangle$ tuples from Reddit comments and 36.5M from submissions. 
They contain 31.4M distinct YouTube videos from 7.5M distinct YouTube channels, mentioned on over 700K distinct subreddits by 1.1M distinct Reddit users.

\xhdr{YouTube data}
Using the YouTube API, we queried the 50 most recent videos for each channel found in our Reddit dataset, obtaining the title, description, category, and the number of views for each video. YouTubers choose the video category among 15 categories, such as Music, Gaming, and Sports. We aggregated the video category channel-wise by majority voting among the categories of each channel's 50 most recent videos.

\xhdr{Recommendations} We retrieved the videos recommended by YouTube for these videos using the InnerTube API,%
which is public and widely used in third-party applications (e.g., see {\url{https://invidious.io/}).
Recommendations provided by the API capture what a history-less user would receive.
Specifically, for each channel, we sample 1 video from its most recent 50 videos with replacement, and then, for each video, query its recommended videos. We repeated this procedure 100 times over four days.

\xhdr{Additional processing}
We reduced spam by removing users who uploaded links to more than $1,000$ videos on Reddit (0.04\% of authors). 
Further, we filtered our data to contain only large channels by restricting to those that have over 100k subscribers (and whose data can be obtained using the YouTube Data API) and that produced their 50 most recent videos mainly in English. 
To determine the language, we resorted to metadata fields, the video caption language, and, if necessary, an additional language detection algorithm.
After this final step, a total of 44,000 channels remain.

\begin{table}
\centering
\begin{tabular}{c|ccc}
\toprule
\textbf{Category} & {t2v-soc} & {t2v-con} & {t2v-rec} \\
\midrule
Gaming & 0.88 & 0.91 & \textbf{0.94} \\
Music & 0.81 & \textbf{0.93} & \textbf{0.94} \\
Sports & 0.86 & 0.90 & \textbf{0.94} \\
\bottomrule
\end{tabular}
\caption{Category prediction: we report the F1 score of random forest classifiers trained to predict the categories using different embeddings as features. Pairwise differences between all scores are statistically significant considering a $z$-test with $\alpha = 0.05$, except for the Music category between the content and recommendation embeddings.}
\label{table_f1_cat}
\end{table}

\section{Channel Embeddings}

\xhdr{Social sharing embedding ({t2v-soc})}
To create a latent representation of how YouTube channels are shared on Reddit, we expand on previous work by \citet{waller_quantifying_2021} and \citet{veselovsky_imagine_2021}.
Specifically, using \citeauthor{waller_quantifying_2021}'s subreddit-level embeddings, represent each YouTube channel vector as a weighted average of the vectors of subreddits where the channel was mentioned, i.e., $C = W \times S$, where the matrix of channel embeddings ($C_{n \times k}$), 
is the product between the matrix of subreddit embeddings ($S_{m \times k}$) and the row-normalized matrix capturing which subreddits shared which channels ($W_{n \times m}$).
Each element $w_{ij} \in W$ represents the number of times a channel $i$ is mentioned on subreddit $j$, divided by the number of times the channel is mentioned on any subreddit.
This approach yielded better embeddings than dimensionality reduction methods on the matrix $W$, e.g., singular value decomposition. 
Further, it allows using  \citeauthor{waller_quantifying_2021}'s ``social dimensions,''  vectors within the subreddit space that were shown to correspond to social constructs, like age and political orientation.

\xhdr{Content embedding ({t2v-con})}
We create a latent representation of the content using video titles and descriptions from each channel's 50 most recent videos. Specifically, we use the \emph{all-MiniLM-L6-v2} pre-trained model adapted from \cite{wang_minilm_2020}. 
We map each video title and description to 384-dimension vectors, which we sum to get a single vector per video.
Then, we obtain a single 384-dimension vector representing each channel by averaging the vectors of its videos.

\xhdr{Recommendation embedding ({t2v-rec})}
Last, we create a latent representation of YouTube recommendations using the recommendation graph we crawled. Specifically, we create a weighted undirected graph where nodes are channels and an edge $\langle u,v, w \rangle$ indicates that $w$ times, a video from a channel $v$ was recommended by a video from channel $u$ or vice versa. 
We only consider nodes and edges associated with channels from our filtered dataset with 44K channels.
With such a graph, we use node2vec to obtain a latent representation for each channel~\cite{grover_node2vec_2016}. We obtained similar results using Deepwalk~\cite{perozzi_deepwalk_2014} and LINE~\cite{tang_line_2015}.

\section{Capturing Semantic Similarity}

\begin{table}
\centering
\begin{tabular}{c|cccc}
\toprule
\textbf{$w$} & {t2v-soc} & {t2v-con} & {t2v-rec} & \textbf{$n$}\\
\midrule
2 & 0.51 & \textbf{0.65} & \textbf{0.68} & 900\\
3 & 0.53 & \textbf{0.67} & \textbf{0.71} &  827\\
4 & 0.59 & \textbf{0.76} & \textbf{0.81} & 544\\
5 & 0.68 & \textbf{0.85} & \textbf{0.89} & 292\\
\bottomrule
\end{tabular}
\caption{Find the `odd-channel-out:'  performance of our classifier for triplets where at least $w$ of workers agreed on the 'odd-channel-out' (e.g., $w=5$ means that all five workers agreed). The differences between content and recommendation embeddings are not statistically significant considering a $z$-test with $\alpha = 0.05$.
}
\label{table_agreement_minwork}
\end{table}

\begin{table}

\begin{tabular}{c|ccc c}
\toprule
$k$ & {t2v-soc} & {t2v-con} & {t2v-rec} & $n$ \\
\midrule
110 & 0.48 & \textbf{0.59} & \textbf{0.60} & 300 \\
220 & 0.53 & \textbf{0.67} & \textbf{0.70} & 300\\
440 & 0.53 & \textbf{0.69} & \textbf{0.73} & 300\\
\bottomrule
\end{tabular}
\centering
\caption{Find the `odd-channel-out:' agreement between workers and embeddings considering different distances ($k$).
The differences between content and recommendation embeddings are not statistically significant considering a $z$-test with $\alpha = 0.05$.}
\label{table_agreement_k}
\end{table}

To determine to which extent different embeddings can capture the semantic similarity of YouTube channels, we use the embeddings to predict the categories of different YouTube channels, and, considering triplets of channels, to predict which channels would be considered `least similar' by human annotators.

\xhdr{Category prediction}
Considering the  ``Music'', ``Sport'', and ``Gaming'' categories (as they are among the ones that most accurately describe the content of channels within them), we trained a Random Forest classifier using the different embeddings.
For each category, we considered 100 channels that produced most of their videos in that category and 100 channels from all other categories and perform a 70/30 train/test split. We repeated each sampling and fit iteration 100 times. 
We report the resulting F1 scores in Table~\ref{table_f1_cat}.
For all categories, the ordering of F1 scores remains the same, with the Social sharing embedding performing the worst (0.85 on average), followed by the content (0.91), and the recommendation embedding (0.94).

\xhdr{Semantic similarity}
We asked crowdworkers on Amazon Mechanical Turk to find the ``odd-channel-out'' from a triplet of 3 YouTube channels, presenting them with the channels' names and the thumbnails and titles of their four most recent videos (see Fig.\ref{fig:mturk}b). For each triplet, we asked five independent crowdworkers to find the  ``odd-channel-out.''
This setup is an adaptation of the word intrusion framework introduced in \cite{NIPS2009_f92586a2} for topic models and applied by \cite{piccardi_crosslingual_2021} for Wikipedia topic models.

Triplets of channels (A, B, C)  were sampled from each embedding such that all three were similar, but one of them is slightly less so.
First, A is sampled at random;  
second, we let B be A's closest neighbor;
and last, C is picked to be A's $k$-th closest neighbor while ensuring that B is closer to A than to C. 
We vary the value $k$ to make our evaluation more robust ($k \in \{110, 220, 440\}$).
We repeat the above procedure  so that an equal number of samples are drawn considering the distance between channels in the social sharing, content, and recommendation embeddings.

In Table~\ref{table_agreement_minwork}, we show the agreement between the embeddings and crowdworkers. 
We report results stratified by the minimum number of crowdworkers (out of five) who agreed on which was the odd channel out. We find that recommendation and content embeddings yield similar results, with the latter obtaining slightly higher scores.
In Table~\ref{table_agreement_k}, we show the agreement between embeddings and crowdworkers with different values of $k$ (the parameter governing the distance of the `odd-channel-out'). We find that the performance of all embeddings is impacted as we lower $k$, but results remain qualitatively the same.

\section{Capturing Social Dimensions}

Here, we consider the social dimensions from \citet{waller_quantifying_2021}, obtained by considering the difference (in the embedding space) between pairs of subreddits that are generally similar but differ in the social dimension of interest, e.g., for the gender dimension, they considered the \texttt{r/Daddit} and \texttt{r/Mommit} communities, which existed to discuss parenting.
For the social sharing embedding, we use the same dimensions as \citet{waller_quantifying_2021} since we have embedded channels in the same latent space as they do.
Then, we extend these social dimensions to the other embeddings by training a Random Forest regressor to predict them with the embedding as the input.
While this requires the social sharing embedding, it enables us to evaluate whether content and recommendation embeddings can capture the social dimensions of YouTube channels.

\xhdr{Pre-existing political labels}
We consider 113 channels labeled as \emph{extreme-left}, \emph{left}, \emph{center-left}, \emph{center}, \emph{center-right}, \emph{right}, \emph{extreme-right} obtained from \citet{mamie_are_2021} and \citet{yoan_dinkov_youtube-political-bias_2018}. 
We compute the rank correlation between the partisan score of each channel and the rank of its label, using Stuart-Kendall rank correlation coefficient $\tau_{c}$, a variant of Kendall's $\tau$ adapted to work with ties \cite{berry_stuarts_2009} (See Table~\ref{dims_table}, first row).
 Overall, the social sharing embeddings perform best ($\tau_{c}=0.67$) followed by recommendation embeddings ($\tau_{c}=0.49$) and content embeddings ($\tau_{c}=0.37$).

\xhdr{Crowdsourced rankings}
To obtain a crowdsourced ranking of channels across social dimensions, we show crowdworkers pairs of channels (again with titles, descriptions, and the most recent videos), and ask them which appeals to someone from one extreme of a social dimension, e.g.,  ``which channel appeals the most to a younger audience?'' (see Fig.~\ref{fig:mturk}a).
With the responses, we build a ranking of channels along that particular dimension with a Plackett-Luce model~\cite{maystre2015fast}; see Appendix~\ref{app:bt}. 

We analyzed three specific channel categories, each paired with a social dimension: ``Music'' for the age dimension, ``News and Politics'' for the partisan dimension, and ``How-to and Style'' for the gender dimension. 
This simplifies the human intelligence task, as comparing semantically similar videos in a social dimension is easier, e.g., comparing a gaming channel to a music channel is harder than comparing two music channels.
To obtain a diverse set of channels, we 
1) standardize the social dimensions; 
2) create 18 bins (which separate channels across the social dimension of interest); and
3) sample 10 channels per bin channels.%
\footnote{To create the bins, we consider the standardized social dimension value of each channel (bin edges:  $\{$$-$5, $-$1.25, $-$0.5, 0.5, 1.25, 5$\}$), but also the -ness dimension associated with each social dimension proposed by \citet{waller_quantifying_2021}. These  capture how salient the social dimension is in each subreddit (bin edges:  $\{$$-$5, 0, 5$\}$); the combination of these two bins yields 18 bins, as mentioned in the main text.}

In Table~\ref{dims_table} (rows 2 to 4), we show the Stuart-Kendall rank correlation coefficient between our crowdsourced ranks and each of our embeddings.
Across all dimensions, the recommendation embedding obtained the highest correlation. Social sharing embeddings performed on par with recommendation embeddings considering the partisan dimension.

\begin{table}
\centering
\begin{tabular}{lrrrr}
\toprule
 & {t2v-soc} & {t2v-con} & {t2v-rec} & $n$\\
\midrule
Pre. Labels & \textbf{0.67} & 0.37 & 0.49 & 113 \\
BT Partisan & \textbf{0.38} & 0.31 & \textbf{0.39} & 100 \\
BT Gender & 0.50 & 0.52 & \textbf{0.58} & 100 \\
BT Age & 0.43 & 0.42 & \textbf{0.51} & 100\\
\bottomrule
\end{tabular}
\caption{We report the Stuart-Kendall rank correlation between different embeddings and 1) pre-existing labels on the political leaning of channels (first row); and 2) crowdsourced social dimension rankings. The differences between social sharing and recommendation embeddings on the  crowdsourced partisan rankings are not statistically significant considering a $z$-test with $\alpha = 0.05$.}
\label{dims_table}
\end{table}

\section{Discussion}

We propose and systematically evaluate a variety of latent representations of YouTube channels.
Using these embeddings, future work could construct datasets of channels through weak supervision or evaluate the prevalence of YouTube content through one or several of the social dimensions considered here.
Further, our analyses suggest that embeddings created through social sharing and recommendation data meaningfully encode social dimensions like partisanship, age, and gender. 
We expect that the embeddings shared with this paper, and the methodology to create and evaluate them, will help computational social scientists study video- and image-centric social media like YouTube.
We advise authors using our embeddings to use the recommendation embeddings provided, except if they are specifically interested in studying partisanship, in which case the social sharing embeddings may be preferred.

Different embeddings each have their limitations. 
Social sharing embeddings rely on Reddit data, and thus we can only embed channels that are shared on another social platform (whose users reside primarily within English-speaking countries). Recommendation embeddings, while the most performant in our analyses, rely on the social sharing dimensions. One could use a methodology similar to \citet{waller_quantifying_2021} to obtain social dimensions in the recommendation space, but empirically, we find that isolating pairs of channels that are similar in all aspects but one is much harder than the analogous subreddit-level task.
Last, content embeddings are perhaps the easiest to obtain but are supplanted by recommendation embeddings in every aspect.

Improvements to our approach could come from many directions.
The recommendation embedding could be improved by gathering recommendations from \textit{innertube} over more extended periods;
the content embedding may improve by using newer models with more oversized context windows (we truncate paragraphs every 256 words);
the social sharing embedding could consider additional social media platforms (like Twitter or Facebook, if data was available for these platforms).
Last, more generally, future work could explicitly model time when creating the embeddings, capturing channels that dramatically change over the years. 

\section*{Ethical Considerations}
We do not foresee a negative societal impact coming from
this research, which, on the contrary, may help researchers to study important social phenomena on platforms like YouTube.
Mechanical Turk workers executing the tasks were paid appropriately to estimate a fair price for our human intelligence tasks; three authors carried out each task to estimate the necessary completion time. We used these to estimate the time necessary to complete the task so that we paid workers 16 dollars an hour.

\appendix
\section{Plackett-Luce}
\label{app:bt}

Considering a social dimension (e.g., age), a given extreme (e.g., old), and two channels $A$ and $B$, let $A  \succ B$ denote that a human rater considers $A$ to be closer to the extreme in the social dimension (e.g., channel $A$ appeals to older individuals than channel $B$).
The Plackett-Luce model assumes that each channel has a latent social dimension score $s$ and that probability that a rater will say that  $A  \succ B$ is proportional to the score $s$:
\begin{equation}
    Pr(A \succ b) = \frac{s(A)}{s(A) + s(B)}
\end{equation}
\citet{maystre2015fast} proposes a maximum-likelihood approach to estimate the latent scores across all channels, which we use to obtain the rankings.

{
\bibliography{bib, references}
}

\end{document}